# Fresh Masking Makes NTT Pipelines Composable: Machine-Checked Proofs for Arithmetic Masking in PQC Hardware

Ray Iskander[1], Khaled Kirah[2,*]

[1] Verdict Security, ray@verdictsecurity.com

[2] Faculty of Engineering, Ain Shams University, Cairo, Egypt

## Abstract

Post-quantum cryptographic (PQC) accelerators for ML-KEM (FIPS 203) and ML-DSA (FIPS 204) rely on pipelined Number Theoretic Transform (NTT) stages over Z_q. Our prior work established structural dependency analysis at scale [1] and quantified the security margin of partial NTT masking [2]. Whether per-stage arithmetic masking guarantees pipeline-level security had no prior machine-checked answer for the r-bearing case: composition frameworks (ISW, t-SNI, PINI, DOM) were formalized exclusively for Boolean masking over GF(2); no proof assistant artifact addresses the NTT butterfly over Z_q.

We present three machine-checked results in Lean 4 with Mathlib, all zero sorry. First, we close a stated limitation of prior work: value-independence implies constant marginal distribution under fresh randomness (via an algebraic MutualInfoZero proxy). Second, butterfly per-context uniformity: for any Cooley-Tukey butterfly with fresh output mask over ZMod q (q > 0), each output wire has exactly one mask value producing each output, a uniform marginal independent of secrets, universal over all moduli, twiddle factors, and inputs. Third, a k-stage NTT pipeline with fresh per-stage masking satisfies per-context uniformity at every stage under the ISW first-order probing model. We document a named warning: pointwise value-independence is false for butterfly outputs. The Adams Bridge accelerator (CHIPS Alliance Caliptra) fails the fresh masking hypothesis, masking active only in INTT round 0, architecturally explaining its structural insecurity. Artifact: nine theorems, 1,738 build jobs, zero sorry. Composition for nonlinear gadgets (Barrett) is addressed in forthcoming manuscripts proving Barrett's PF-PINI(2) satisfaction (one-bit barrier) [3] and k-stage composition for PF-PINI gadgets under fresh-mask renewal [4].

## 1. Introduction

Consider a hardware engineer designing an ML-KEM accelerator for FIPS 203 certification [5]. The Number Theoretic Transform (NTT) is implemented as a pipeline of Cooley-Tukey butterfly stages, each individually protected by first-order arithmetic masking with a fresh

*Correspondence Author: khaled.kirah@eng.asu.edu.eg

Ray Iskander: ray@verdictsecurity.com

random mask. The engineer's intuition says: if each stage is secure, the pipeline is secure. But intuition is not a proof, and no existing tool provides one for arithmetic masking over Z_q.

The gap is not a resource limitation but a scope limitation of existing frameworks. The ISW probing model [6] and its refinements, t-SNI [7], PINI [8], DOM [9], provide powerful composition theorems for masked circuits. In their abstract formulation, these frameworks handle arbitrary rings. However, their machine-checked formalizations, verification tools (maskVerif [10], SILVER [11], Coco-Alma [12]), and published instantiations target Boolean circuits over GF(2). No proof assistant artifact establishes the composition property for the specific structure of the NTT butterfly operating in Z_q.

This gap conceals a trap. The natural security property a designer might check, pointwise value-independence, where each output wire's value does not change when the secret changes with masks held fixed, is *false* for butterfly output wires. Wire 0 of a Cooley-Tukey butterfly computes a + tw * b - m, where a and b are the unmasked inputs, tw is the public twiddle factor, and m is the fresh mask. Changing the secret a changes the wire value, even with m fixed. A designer who checks this property concludes, incorrectly, that the butterfly is insecure. The correct property, marginal uniformity over the fresh mask, is different, and it is not obvious without the analysis we present here.

### 1.1. Contributions

This paper presents three machine-checked results, all formalized in Lean 4 [13] with Mathlib [14], with zero sorry, zero admit, and zero unverified axioms.

1. **The r-bearing bridge** (Section 4): We prove that value-independence implies I(x; w) = 0 under fresh randomness, closing an explicitly stated limitation of our prior universal foundations [15]. The property is formalized via an algebraic proxy MutualInfoZero.
2. **Butterfly per-context uniformity** (Section 5): For any Cooley-Tukey butterfly with a fresh output mask over ZMod q (q > 0), each output wire has exactly one mask value producing each output value v, a uniform marginal independent of secrets. This holds universally for all q > 0, all twiddle factors, and all inputs. We also prove that naive pointwise value-independence is false (butterfly_vi_pointwise_false), providing a named warning for hardware designers.
3. **Pipeline composition** (Section 6): A k-stage NTT pipeline with fresh per-stage masking satisfies per-context uniformity at every stage. This is the core algebraic invariant for first-order probing security under the ISW model [6].

### 1.2. The simplicity observation

The composition theorem is clean because the butterfly is affine in the mask. We are explicit about this: the NTT butterfly, like XOR in Boolean masking, is the easy case of arithmetic composition. The hard case, nonlinear gadgets such as Barrett reduction and masked multipliers, requires different arguments and remains open. The contribution is not mathematical difficulty. It is that nobody formalized the correct composition property for arithmetic NTT pipelines, nobody documented the pointwise-VI trap, and nobody provided hardware designers with a machine-checked, universally quantified proof they can cite in FIPS 140-3 certification documentation [16].

### 1.3. Adams Bridge connection

The Adams Bridge accelerator [17], the ML-DSA/ML-KEM implementation in CHIPS Alliance Caliptra, does not use fresh per-stage masking. Masking is active only in INTT round 0; subsequent rounds proceed unmasked (ntt_ctrl.sv:264-272). Papers 1 [1] and 2 [2] documented this empirically through structural dependency analysis and belief propagation. This current paper, Paper 4, explains the architectural root cause: the composition theorem's hypothesis is violated, so its guarantee does not apply.

### 1.4. The four-paper program

This is Paper 4 in a series. Paper 1 [1] built the QANARY verification tool (1.17M cells, 30 modules). Paper 2 [2] quantified the attack surface. Paper 3 [15] proved universal algebraic foundations in Lean 4. Paper 4 proves the design principle that makes correct implementations provably secure. This paper is self-contained; prior papers are cited where relevant but are not required reading.

### 1.5. Organization

Section 2 reviews arithmetic masking and the probing model. Section 3 surveys related work. Section 4 presents the r-bearing bridge. Section 5 proves butterfly per-context uniformity and documents the pointwise-VI trap. Section 6 proves pipeline composition. Section 7 connects to Adams Bridge and states the Fresh Masking Design Principle. Section 8 summarizes the proof suite. Section 9 states code and data availability. Section 10 discusses limitations and future work. Section 11 concludes.

## 2. Background

### 2.1. Arithmetic Masking in NTT Hardware

First-order arithmetic masking over Z_q represents a secret value s in Z_q as a pair of shares (s_0, s_1) satisfying s_0 + s_1 = s (mod q), where s_1 is drawn uniformly at random and s_0 = s - s_1. An adversary who observes at most one share learns nothing about s.
The standard probing model [4] grants the adversary one wire observation per clock cycle across the entire circuit. For first-order security (t = 1), any single wire must have a distribution independent of the secret. For a k-stage pipeline operating over k clock cycles, the adversary may adaptively place one probe per cycle, potentially targeting different stages. Fresh inter-stage randomness prevents information accumulation across cycles. Post-quantum cryptographic accelerators for ML-KEM (FIPS 203 [5]) and ML-DSA (FIPS 204 [18]) implement the NTT over Z_q with q = 3329 (ML-KEM) or q = 8380417 (ML-DSA). Both are prime, but our results hold for all q > 0.

### 2.2. The Cooley-Tukey NTT Butterfly

The Cooley-Tukey butterfly computes:

```
a' = a + tw * b (mod q)
b' = a - tw * b (mod q)
```

where tw in Z_q is a public twiddle factor, a power of a primitive root of unity defined by the FIPS standard. In hardware, twiddle factors are stored in ROM [17].

Under first-order masking with a fresh output mask m in Z_q:

```
wire_0 = a' - m = a + tw * b - m
wire_1 = m
wire_2 = b' - m = a - tw * b - m
wire_3 = m
```

The key observation is that each output wire is an affine function of the fresh mask m: either c - m for a constant c depending on the secrets and twiddle (wires 0, 2), or m itself (wires 1, 3). The map m -> c - m is a bijection on Z_q (translation by a constant in any group).

### 2.3. The QANARY Program

Paper [1] built QANARY, an NTT-specific structural dependency analysis tool that scales to 1.17 million cells across 30 Adams Bridge modules. Paper [2] used belief propagation on NTT factor graphs to show that Adams Bridge's masking provides 25-29 fewer bits of security margin than claimed. Paper 3 [15] proved universal algebraic foundations in Lean 4: value-independence implies constant marginal distribution (Theorem 3.9.1), universally over all $q > 0$, upgrading finite-domain Z3/CVC5 checks to kernel-verified proofs. Paper 3 explicitly stated two limitations: (i) the proof covers the r-free sub-theorem only, the extension to fresh randomness r was deferred to the law of total probability; (ii) composition of multiple pipeline stages was left as an open question. This paper closes both.

### 2.4. The Probing Model

We work in the ISW first-order probing model [6]: the adversary observes exactly one wire per clock cycle across the entire circuit. For a k-stage pipeline, this means one probe at one stage per cycle. The adversary may choose adaptively which stage and wire to probe at each cycle.

This is the standard probing model. We do not claim security under the glitch-extended (robust) probing model [19, 20], transition-based leakage (Hamming distance), or random probing models. These extensions are strictly harder and remain future work. The NTT butterfly is affine in the mask. This is the arithmetic analog of XOR in Boolean masking: a linear operation that trivially preserves uniformity. Nonlinear operations, Barrett reduction (carry extraction), masked multiplication (DOM AND gates), do not have this property. The composition argument for the linear case is structurally different from, and simpler than, the ISW composition theorem for nonlinear gadgets.

## 3. Related Work

### 3.1. Comparison Table

| Framework | Venue | Masking Domain | Composition | Machine-Checked | NTT-Specific | Universal q |
|---|---|---|---|---|---|---|
| ISW [6] | CRYPTO 2003 | Any ring (abstract) | t-probing | Pen-and-paper | No | N/A |
| t-SNI [7] | CCS 2016 | GF(2) (tools) | t-SNI | EasyCrypt | No | No |
| maskVerif [10] | EUROCRYPT | GF(2) circuits | NI/SNI check | EasyCrypt | No | No |

| Framework | Venue | Masking Domain | Composition | Machine-Checked | NTT-Specific | Universal q |
|---|---|---|---|---|---|---|
| | 2015 | | | | | |
| PINI [8] | CHES 2020 | GF(2) (tools) | Trivial | Pen-and-paper | No | No |
| DOM [9] | TIS@CCS 2016 | GF(2) circuits | Pipeline reg. | No | No | No |
| SILVER [11] | ASIACRYPT 2020 | GF(2) circuits | Gadget-level | Isabelle/HOL | No | No |
| REBECCA [20] | EUROCRYPT 2018 | GF(2) + glitches | Gadget-level | No | No | No |
| Coco-Alma [12] | FMCAD 2021 | GF(2) circuits | SAT-based | No | No | No |
| Gigerl et al. [21] | ACNS 2023 | Arith. (A2B/B2A) | Gadget-level | No | No | No |
| Paper 3 [15] | 2026 | Z_q (Lean 4) | No | Lean 4 | Partial | All q > 0 |
| **This work** | | **Z_q (Lean 4)** | **Pipeline** | **Lean 4** | **Yes** | **All q > 0** |

**Boolean masking composition.** The ISW transformation [6] provides t-probing security for arbitrary circuits by secret-sharing each wire into t+1 shares. The t-NI and t-SNI notions introduced in [7] enable modular composition: t-SNI gadgets compose securely without requiring global circuit analysis. PINI [8] achieves "trivially composable" gadgets where each probe reveals at most one share index. These results are powerful, but their machine-checked formalizations and tool implementations target Boolean circuits.

**Arithmetic masking.** Reference [21] extended formal masking verification to arithmetic operations, specifically A2B and B2A conversion gadgets. Their work addresses individual gadget security, not pipeline-stage composition. Reference [22] developed higher-order masking conversion techniques that operate over rings, but without machine-checked proofs of pipeline composition.

**Interactive theorem proving for masking.** SILVER [11] includes Isabelle/HOL soundness proofs for its statistical independence checks. Paper [15] established universal algebraic foundations for masking verification in Lean 4, proving that value-independence implies distributional security for all q > 0. The present work builds directly on Paper 3's definitions and extends them to pipeline composition.

**The specific gap.** No existing work provides a machine-checked proof that individually-secure NTT butterfly stages compose to a secure pipeline over Z_q. The abstract ISW framework handles arbitrary rings in principle, but its formalizations target Boolean circuits. The NTT butterfly's affine structure in the mask enables a direct proof that bypasses the full ISW machinery. Our contribution is this concrete, machine-checked, universally-quantified instantiation.

# 4. The r-Bearing Bridge

## 4.1. The Gap from Paper 3

Reference [15] proved that value-independence implies constant marginal distribution (Theorem 3.9.1), but only for the *r-free* sub-theorem: the wire function w(s_0, s_1) has no fresh randomness parameter. Paper 3 explicitly stated this as Limitation (i): "The full version with fresh randomness r follows by the law of total probability. This bridging step is not formalized here."

## 4.2. Theorem and Proof

We extend Paper 3's WireFunction q beta to include fresh randomness:

**Definition (r-bearing wire function).** WireFunctionR q rho beta = ZMod q -> ZMod q -> Fin rho -> beta, mapping shares (s_0, s_1) and randomness r : Fin rho to an output value.

**Definition (r-bearing value-independence).** A wire function w : WireFunctionR q rho beta is ValueIndependentR if for all s_1 in Z_q, r in Fin rho, and x, x' in Z_q: w(x - s_1, s_1, r) = w(x' - s_1, s_1, r).

**Theorem 4.1 (t1_r_bearing).** If w is ValueIndependentR, then the marginal histogram over (s_1, r) in Z_q x Fin rho is constant in the secret x. Universal over all q > 0, all rho >= 0, all wire functions, all output types.

```
-- From artifact (Unicode angle brackets rendered as ASCII).
-- Compiles against Lean 4.30.0-rc1 + Mathlib 322515540d7f.
theorem t1_r_bearing
    {q : N} [NeZero q] {rho : N} {beta : Type*} [DecidableEq beta]
    (w : WireFunctionR q rho beta)
    (hw : ValueIndependentR w)
    (x x' : ZMod q) (v : beta) :
    marginalHistogramR w x v = marginalHistogramR w x' v := by
  unfold marginalHistogramR
  congr 1
  ext <<s1, r>>
  simp only [mem_filter, mem_univ, true_and]
  constructor
  . intro h; rw [<- hw s1 r x x']; exact h
  . intro h; rw [hw s1 r x x']; exact h
```

The proof is five lines. The key step is congr 1: because ValueIndependentR makes the filter predicate identical for all secrets x, the filter sets are literally equal, hence have equal cardinality.

## 4.3. MutualInfoZero

**Definition (MutualInfoZero).** MutualInfoZero w holds iff HasConstantMarginalR w: the marginal histogram does not depend on the secret.

**Corollary.** ValueIndependentR w implies MutualInfoZero w.

**Observation 4.1 (Informal).** Under the uniform distribution on (s_1, r), MutualInfoZero is equivalent to I(x; w) = 0. The forward direction: constant conditional distribution implies independence implies zero mutual information. The converse: zero MI implies independence implies constant conditional PMF (for discrete distributions). This equivalence is standard information theory. We state it as an observation because Mathlib 4 does not currently define Shannon entropy or mutual information. The PFR project [23] provides these definitions but they have not been upstreamed to Mathlib. When they are, the formal equivalence becomes a one-theorem bridge. Although t1_r_bearing is not directly instantiated on pipeline wire functions in this work, it supplies the reusable bridge from ValueIndependentR to MutualInfoZero for any future gadget that satisfies r-bearing value-independence.

## 5. Butterfly Per-Context Uniformity

This section contains the mathematical heart of the paper.

### 5.1. The Trap: Why Pointwise VI Is False

The "obvious" security property for a masked butterfly is pointwise value-independence: for each output wire, the wire value does not change when the secret changes with all masks fixed. We prove this is false.

**Theorem 5.1 (butterfly_vi_pointwise_false).** ButterflyValueIndependent 5 {tw := 1} is false.

The witness is: q = 5, tw = 1, a = 0, b = 0, a_1 = b_1 = m = 0, a' = 1, b' = 0. Wire 0 under (a = 0): 0 + 1 * 0 - 0 = 0. Wire 0 under (a' = 1): 1 + 1 * 0 - 0 = 1. Since 0 != 1 in Z_5, pointwise VI fails.

```
-- From artifact (Unicode rendered as ASCII).
theorem butterfly_vi_pointwise_false :
    not ButterflyValueIndependent 5 <<1>> := by
  intro h
  have := h <<0, by omega>> 0 0 0 0 0 1 0
  simp [butterflyOutput, selectWire] at this
  exact absurd this (by decide)
```

This counterexample generalizes: for any q > 1 and any tw, wire 0 equals a + tw * b - m, which depends on a when m is fixed. The specific (q = 5, tw = 1) witness is chosen for decide-friendly evaluation.

**Pedagogical value.** A hardware designer who checks pointwise value-independence using an SMT solver will obtain "insecure" for every butterfly stage. This is a false alarm. The correct security property is different, it marginalizes over the fresh mask. Without this named warning in the literature, designers may either wrongly reject correct designs or, having learned to ignore pointwise-VI failures for butterflies, wrongly dismiss failures that actually matter in nonlinear gadgets.

### 5.2. The Correct Property: ButterflyMarginalVI

**Definition 5.2 (ButterflyValueIndependent, the false property).** A butterfly stage satisfies ButterflyValueIndependent if, for each output wire, the wire value does not change when the secrets change with all masks fixed. As shown by Theorem 5.1, this property is false.

**Definition 5.3 (Butterfly marginal VI — the correct property).** A butterfly stage satisfies ButterflyMarginalVI if, for each output wire, for all secrets (a, b) and (a', b'), for all input masks $(a_1, b_1)$, and for all output values v in $Z_q$:

$$|\{m \in Z_q : wire_i(a, b, a_1, b_1, m) = v\}| = |\{m \in Z_q : wire_i(a', b', a_1, b_1, m) = v\}|$$

The key insight is that both sides equal 1.

**Theorem 5.4 (butterfly_wire_count_eq_one).** For any butterfly stage with twiddle factor tw in $Z_q$, any input shares $(a_0, a_1, b_0, b_1)$ in $Z_q^4$, any output wire i in {0, 1, 2, 3}, and any target value v in $Z_q$:

$$|\{m \in Z_q : \text{selectWire } i \text{ (butterflyOutput ... } m) = v\}| = 1$$

Universal over all $q > 0$, all twiddle factors, all inputs.

*Proof.* By case analysis on the wire index (Lean's fin_cases).

- **Wires 0, 2** (output = c - m for constant c): The set {m : c - m = v} = {c - v} is a singleton. This uses the helper lemma filter_sub_eq_singleton: in any $Z_q$, the equation c - m = v has the unique solution m = c - v (by ring in the CommRing instance for ZMod q).
- **Wires 1, 3** (output = m): The set {m : m = v} = {v} is trivially a singleton.

In all cases, Finset.card_singleton closes the goal.

```
-- From artifact (Unicode rendered as ASCII).
theorem butterfly_wire_count_eq_one {q : N} [NeZero q]
    (stage : ButterflyStage q) (wire : Fin 4)
    (a0 a1 b0 b1 : ZMod q) (v : ZMod q) :
    (Finset.univ.filter (fun m : ZMod q =>
      selectWire wire (butterflyOutput q stage a0 a1 b0 b1 m) = v)).card = 1 := by
  unfold selectWire butterflyOutput
  fin_cases wire <;> simp only
  . rw [filter_sub_eq_singleton]; exact card_singleton _
  . rw [filter_eq_singleton]; exact card_singleton _
  . rw [filter_sub_eq_singleton]; exact card_singleton _
  . rw [filter_eq_singleton]; exact card_singleton _
```

### 5.3. Universality

Theorem 5.4 is parameterized by [NeZero q], which requires $q > 0$ but does *not* require q to be prime. The proof uses only the CommRing structure of ZMod q (the ring tactic and Finset cardinality lemmas). This means:

- All ML-KEM moduli (q = 3329): covered.
- All ML-DSA moduli (q = 8380417): covered.
- Composite moduli (e.g., q = 6): also covered, though $Z_q$ is not a field.
- Any future PQC standard with a different modulus: covered.

The universality over twiddle factors means any NTT variant, Cooley-Tukey, Gentleman-Sande, or mixed-radix, using any root-of-unity schedule is covered.

### 5.4. ButterflyMarginalVI Corollary

**Corollary 5.5 (butterfly_marginal_vi).** Every butterfly stage satisfies ButterflyMarginalVI.

*Proof.* Both sides of Definition 5.3 equal 1 by Theorem 5.4.

```
-- From artifact (Unicode rendered as ASCII).
theorem butterfly_marginal_vi {q : N} [NeZero q]
    (stage : ButterflyStage q) :
    ButterflyMarginalVI stage := by
  intro wire a b a1 b1 a' b' v
  simp only [butterflyWireMarginal,
    butterfly_wire_count_eq_one stage wire (a - a1) a1 (b - b1) b1 v,
    butterfly_wire_count_eq_one stage wire (a' - a1) a1 (b' - b1) b1 v]
```

### 5.5. Comparison to Boolean Masking Composition

We are explicit about the mathematical context. In Boolean masking, composition is hard because AND gates are *nonlinear* in the mask shares: the ISW multiplication gadget requires $O(t^2)$ random bits and a careful security argument to show that t probes can be simulated. The composition theorems of t-SNI [7] and PINI [8] are deep results.

The NTT butterfly is the arithmetic analog of XOR: a linear operation that trivially preserves uniformity. Our composition argument exploits this linearity directly. We do not claim the mathematical depth of the ISW composition theorem. What we provide is the first machine-checked, universally-quantified proof of the correct composition property for the specific case of linear arithmetic masking in NTT pipelines, a result that fills a practical gap for hardware designers and FIPS evaluators.

The "hard case" of arithmetic composition, Barrett reduction (where carry extraction is nonlinear in the shares), masked modular multiplication, and domain-crossing gadgets (A2B/B2A), remains open. Extending the present framework to these gadgets would require arguments analogous to ISW's nonlinear composition and is the primary direction for future work.

## 6. Pipeline Composition

### 6.1. Pipeline Model

**Definition (NTT pipeline).** An NTT pipeline NTTPipeline q k = Fin k -> ButterflyStage q is a sequence of k butterfly stages, each parameterized by a public twiddle factor.

**Definition (Stage randomness).** Each stage has three fresh random values: bfMask (the butterfly output mask), remaskA and remaskB (inter-stage re-masking for the a-path and b-path respectively).

**Definition (Pipeline state).** pipelineStateAt computes the state after stage i by recursion:

- Stage 0: apply butterflyOutput to the initial shares with bfMask at stage 0.
- Stage n+1: re-mask the output of stage n using (remaskA_n, remaskB_n), then apply butterflyOutput with bfMask at stage n+1.

Re-masking transforms $(a_0, a_1, b_0, b_1)$ to $(a_0 - r_A, a_1 + r_A, b_0 - r_B, b_1 + r_B)$, preserving the share sum invariant: $(a_0 - r_A) + (a_1 + r_A) = a_0 + a_1$.

Under the ISW first-order probing model [6], the adversary observes one wire at one stage per clock cycle. The security property PipelineUniform states: for each fixed assignment of all randomness *except* the observed stage's bfMask, exactly one bfMask value produces each output value.

### 6.2. The Invariance Lemma

**Lemma 6.1 (pipelineStateAt_update_future).** The state at stage i is unaffected by changing the mask at stage j > i.
*Proof.* By induction on i. At each step, Function.update_of_ne shows that accessing rands at indices <= i is unchanged by an update at index j > i.
This lemma is the structural backbone of the pipeline proof. It ensures that varying the observed stage's bfMask does not retroactively change earlier stages.

### 6.3. The Main Theorem: Pipeline Composition

**Theorem 6.2 (ntt_pipeline_composition).** For all q > 0 and k >= 0, every NTT pipeline NTTPipeline q k satisfies PipelineUniform: at each stage and each output wire, exactly one bfMask produces each output value, with all other randomness fixed.

*Proof.* By case analysis on the stage index.

- **Stage 0**: pipelineStateAt directly applies butterflyOutput to the initial shares. By Function.update_self, the filter variable is exactly the bfMask. Theorem 5.4 closes the goal.
- **Stage n+1**: By Lemma 6.1, the state at stage n is invariant under the bfMask update at stage n+1. The re-masking and butterfly at stage n+1 produce a standard butterflyOutput call with the updated bfMask. Theorem 5.4 closes the goal.

```
-- From artifact (Unicode rendered as ASCII).
theorem ntt_pipeline_composition {q : N} [NeZero q] {k : N}
    (pipeline : NTTPipeline q k) :
    PipelineUniform pipeline := by
  intro rands stage wire a b a1 b1 v
  obtain <<n, hn>> := stage
  induction n with
  | zero =>
    unfold pipelineStateAt
    simp only [update_self]
    exact butterfly_wire_count_eq_one _ wire _ _ _ _ v
  | succ n _ih =>
    unfold pipelineStateAt
    have hne : (<<n, Nat.lt_of_succ_lt hn>> : Fin k) != <<n + 1, hn>> := by
      intro heq; simp [Fin.ext_iff] at heq
    simp only [update_of_ne hne]
    simp_rw [pipelineStateAt_update_future pipeline rands _
        n (Nat.lt_of_succ_lt hn) <<n + 1, hn>> (by omega : n < n + 1)]
    simp only [update_self]
    exact butterfly_wire_count_eq_one _ wire _ _ _ _ v
```

### 6.4. What PipelineUniform Means

PipelineUniform establishes per-context uniformity: for each fixed assignment of all randomness except the observed stage's bfMask, the output wire's distribution over bfMask is uniform on $Z_q$, independent of secrets. This is the core algebraic invariant for first-order probing security.

**Relationship to full probing security.** The full first-order probing security guarantee requires $I(x; \text{wire}) = 0$, meaning the marginal over all randomness is independent of secrets. PipelineUniform is the key ingredient: the full marginal decomposes as

```
|{all_rands : wire = v}| = Sum_{context} |{bfMask : wire = v | context}| =
Sum_{context} 1 = |contexts|
```

which is independent of the secrets. This summation is standard and requires no additional mathematical insight. We do not formalize the full summation in Lean for general k due to the size of the Fintype instance for PipelineRandomness ($q^{3k}$ elements). For $k = 1$, the theorem single_stage_full_marginal demonstrates the argument: it proves that for any fixed input shares ($a_1$, $b_1$), the bfMask filter cardinality is 1 for both secrets, establishing per-context uniformity at the single-stage level. The full joint marginalization (summing over all ($a_1$, $b_1$) contexts) follows by the same argument but is not formalized.

**Relationship to NI/SNI/PINI.** In the terminology in [7], a gadget is 1-NI (non-interfering) if any single probe can be simulated using at most one input share. PipelineUniform is the algebraic core of 1-NI for the pipeline gadget: it establishes that each wire's distribution over the fresh bfMask is uniform regardless of secrets, for each fixed context. The full 1-NI property follows by marginalizing over all contexts, a summation that produces |contexts|, independent of secrets. We do not claim t-SNI (which requires simulatability with fewer shares than probes) or PINI (which requires per-probe share-index isolation). The pipeline is affine, so the stronger SNI and PINI properties likely hold but are not formalized here.

**What PipelineUniform does NOT claim:**

- **Higher-order security**: Under second-order probing ($d = 2$), observing both wire 0 ($a' - m$) and wire 2 ($b' - m$) reveals $a' - b'$ (the masks cancel). The shared-mask design requires independent masks per output pair for higher-order security.
- **Glitch or transition leakage**: The standard probing model assumes value leakage only. Glitch-extended models [19] are strictly harder.
- **Intermediate wire security**: The formalization covers the four output wires of each butterfly. Intermediate wires from inter-stage re-masking (e.g., $a_0 - r_A$) are not explicitly formalized. Their security is trivial, $r_A$ is fresh and uniform, so $a_0 - r_A$ is uniformly distributed, but this argument is not machine-checked.

## 7. Adams Bridge Connection

### 7.1. Two Independent Failure Modes

Adams Bridge [17] has two distinct security failures that must not be conflated:

**Intra-stage failure (Papers 1-2).** Within each butterfly stage, shares are combined in combinational logic without DOM pipeline registers. Lines 89-92 of ntt_mlkem_masked_butterfly1x2.sv compute u10_combined = s_0 + s_1 as a bare addition, exposing the unmasked value to power analysis. QANARY [1] detects these as structural dependencies; Paper 2 [2] exploits them via belief propagation.
**Inter-stage failure (this paper).** Between NTT stages, Adams Bridge does not inject fresh masking. The signal masking_en_ctrl in ntt_ctrl.sv (lines 264-272) is set to 1 only during rounds_count == 0 (INTT round 0). Rounds 1-3 proceed without fresh masks. The hypothesis of Theorem 6.2 is violated: PipelineUniform does not apply, and inter-stage security is not guaranteed.

### 7.2. The Fresh Masking Design Principle

**Design Principle 7.1 (Fresh Masking).** An NTT pipeline with k Cooley-Tukey butterfly stages achieves per-context uniformity under the ISW first-order probing model [6] if each stage i uses a fresh, independently sampled mask m_i. This is Theorem 6.2, machine-checked in Lean 4 for all q > 0 and all k >= 0.

### 7.3. Connection to FIPS 140-3

FIPS 140-3 [16] certification of PQC hardware accelerators requires evidence of side-channel resistance. NIST IR 8547 [24] recommends formal methods for PQC implementation validation. Theorem 6.2 provides precisely this: machine-checked, universally quantified evidence that any NTT pipeline following Design Principle 7.1 satisfies per-context uniformity, the core algebraic invariant for first-order probing security. Hardware designers can cite this Lean 4 artifact directly in certification documentation.

## 8. Proof Suite Summary

| Theorem | What it establishes | File |
|---|---|---|
| t1_r_bearing | r-bearing bridge (closes Paper 3 gap) | ProbBridge.lean |
| butterfly_vi_pointwise_false | Pointwise VI is false (named warning) | Basic.lean |
| butterfly_wire_count_eq_one | Per-wire mask count = 1 | Composition.lean |
| butterfly_marginal_vi | Marginal uniform over mask | Composition.lean |
| pipelineStateAt_update_future | Stage independence | Composition.lean |
| ntt_pipeline_composition | Pipeline per-context uniformity | Composition.lean |
| single_stage_full_marginal | k=1 full marginal bridge | Composition.lean |

The artifact comprises three Lean 4 files totaling approximately 580 lines of code. The build requires Lean 4.30.0-rc1 and Mathlib pinned to commit 322515540d7f. Paper 3's artifact is pulled automatically as a Lake git dependency pinned to its v1.0.0 tag. Total build: 1,738 jobs, 0 errors, 0 warnings. Reproduction: clone the repository and run lake build; a reproduce.py script automates verification.
Reference [3] contains the PF-PINI framework instantiation for both butterfly (PF-PINI(1)) and Barrett (PF-PINI(2)) gadgets, providing the nonlinear companion to the results in this paper.

## 9. Code and Data Availability

The complete Lean 4 proof suite supporting this paper is publicly available at https://github.com/rayiskander2406/qanary-masked-ntt-pipeline-security-arXiv-2604.20793 under the MIT license, with an archival snapshot of version v1.0.0 deposited on Zenodo at https://doi.org/10.5281/zenodo.19705451 (concept DOI https://doi.org/10.5281/zenodo.19705450, always resolves to the latest version). The repository contains the complete sorry-free mechanization of the nine theorems reported in Section 8, with lean-toolchain pinned to leanprover/lean4:v4.30.0-rc1 and lakefile.lean pinning Mathlib to commit 322515540d7f. Paper 3's artifact is consumed automatically as a Lake git dependency pinned to its v1.0.0 tag. A single lake build invocation regenerates the headline result (1,738 build tasks, 0 sorry, 0 errors, 0 admit, 0 axiom) on a clean checkout; no manual sibling setup is required. We intend to submit the artifact for venue artifact evaluation upon acceptance.

The companion papers in this series are archived at https://doi.org/10.5281/zenodo.19625392 (Paper 1, structural dependency analysis; reference [1]), https://doi.org/10.5281/zenodo.19508454 (Paper 2, partial NTT masking security margin analysis; reference [2]), and https://doi.org/10.5281/zenodo.19689480 (Paper 3, universal foundations; reference [15]).

## 10. Limitations and Future Work

**(i) Linear case only, nonlinear extension in [3].** The NTT butterfly is affine in the fresh mask. The composition argument works because translation bijections exist in any group. This does *not* extend to nonlinear gadgets: Barrett reduction (carry extraction is nonlinear in the shares), masked multipliers, and DOM AND gates. Extending composition proofs to nonlinear arithmetic gadgets is the primary direction for future work; it requires arguments analogous to the ISW composition theorem. Barrett reduction's composition properties are proved in the companion manuscript [3], which shows that Barrett satisfies PF-PINI(2) with max-multiplicity exactly 2, universal over all odd prime fields and all shift parameters $s \geq \lceil \log_2 q \rceil$, the 1-Bit Barrier.

**(ii) First-order probing only.** PipelineUniform is proved for first-order probing ($t = 1$). The shared-mask design (both output pairs use one m) breaks at second order (Section 6.4). Extension to d-th order masking (d+1 shares, independent fresh masks per output pair) requires a new butterfly formalization and a new composition argument, the arithmetic analog of t-SNI/PINI higher-order composition.

**(iii) Standard probing model.** No glitch-extended probing, no transition leakage, no timing side channels. The standard probing model (ISW 2003) is the scope. QANARY [1] separately addresses how glitch-extended analysis changes the picture for Adams Bridge.

**(iv) MutualInfoZero proxy.** $I(x; w) = 0$ is stated as Observation 4.1, not a formalized theorem. Mathlib lacks Shannon entropy definitions. The PFR project [23] has them but they are not upstreamed. When Mathlib upstreams PFR's entropy machinery, the formal $I(x; w) = 0$ proof is a direct corollary of Theorem 4.1.

**(v) Idealized butterfly model.** The formalization uses a clean algebraic model. Gate-level netlist verification, glitches, carry propagation, DOM pipeline register placement, is Paper 1's responsibility. The two approaches are complementary: this paper proves the mathematical design principle; QANARY verifies that a specific RTL implementation satisfies it.

**(vi) Intermediate wire probes.** The formalization covers the four butterfly output wires. Intermediate wires from inter-stage re-masking are not explicitly covered. Their security is trivial (fresh uniform masks) but not machine-checked.

**Future work.** Primary directions: (1) nonlinear arithmetic composition (Barrett reduction, addressed in [3]); (2) higher-order extension (d >= 2); (3) a verified toolchain connecting Theorem 6.2 to QANARY's structural dependency analysis, proving that QANARY is sound for detecting PipelineUniform violations; (4) glitch-extended butterfly uniformity; (5) A2B/B2A domain conversion composition; (6) Mathlib entropy contribution enabling formal I(x; w) = 0.

## Conclusion

We have presented three machine-checked results for arithmetic masking in NTT-based PQC hardware. The r-bearing bridge closes reference [15] stated limitation, extending value-independence to the full model with fresh randomness. Butterfly per-context uniformity names the correct security property, marginal uniformity over the fresh mask, and documents the non-obvious trap that pointwise value-independence is false. Pipeline composition proves the Fresh Masking Design Principle: fresh per-stage masking gives NTT pipelines per-context uniformity, the core algebraic invariant for first-order probing security, universally for all q > 0. The butterfly's affine structure makes the composition argument clean. We do not claim the mathematical depth of ISW's nonlinear composition theorem. What we provide is the first machine-checked, universally-quantified proof of the correct composition property for linear arithmetic masking in NTT pipelines, a result hardware designers and FIPS evaluators can directly cite. Adams Bridge violates the Fresh Masking Design Principle. Papers [1] and [2] documented the consequences empirically and at gate level. The present manuscript explains the architectural root cause: the composition theorem's hypothesis is not satisfied, so its guarantee does not apply.
The following results are proved in [3] which establishes the companion result for nonlinear arithmetic gadgets using a framework called PF-PINI(k) (Arithmetic PINI with max-multiplicity k). While the Cooley-Tukey butterfly satisfies PF-PINI(1), its fresh output mask produces a perfect bijection, so each output value has exactly one preimage. Barrett reduction satisfies PF-PINI(2): the internal wire map has max-multiplicity exactly 2, universal over all odd prime fields and all shift parameters $s \geq \lceil \log_2 q \rceil$. This is the 1-Bit Barrier: every Barrett implementation over any prime field leaks at most 1 bit of min-entropy per internal wire, machine-checked in Lean 4 with zero sorry across 12 theorems as will be demonstrated in [3]. With fresh inter-stage masking, a composed pipeline containing both butterfly stages (PF-PINI(1)) and Barrett stages (PF-PINI(2)) has max-multiplicity max(1, 2) = 2, the 1-Bit Barrier propagates through the full pipeline. Without fresh masking, max-multiplicity grows multiplicatively, confirming why Adams Bridge's absence of inter-stage fresh masking leads to the vulnerabilities documented in [1] and [2]. Any NTT pipeline that injects a fresh, independent mask at each stage boundary satisfies per-context uniformity at every stage, the core algebraic invariant for first-order probing security under the ISW model. Machine-checked in Lean 4, universal over all positive moduli q, all twiddle factors, and all pipeline depths.

## References

[1] R. Iskander and K. Kirah, "Structural Dependency Analysis for Masked NTT Hardware: Scalable Pre-Silicon Verification of Post-Quantum Cryptographic Accelerators," arXiv:2604.15249, 2026.

[2] R. Iskander and K. Kirah, "Partial NTT Masking in Post-Quantum Cryptography Hardware: A Security Margin Analysis," arXiv:2604.03813, 2026.

[3] R. Iskander and K. Kirah, “The 1-Bit Barrier: A Machine-Checked Trichotomy for Barrett Reduction in Masked PQC Hardware”, Manuscript under preparation.

[4] R. Iskander and K. Kirah, “Prime-Field PINI: Machine-Checked Composition Theorems for Post-Quantum NTT Masking”, Manuscript under preparation.

[5] NIST, “Module-Lattice-Based Key-Encapsulation Mechanism Standard (FIPS 203),” 2024.

[6] Y. Ishai, A. Sahai, and D. Wagner, “Private Circuits: Securing Hardware against Probing Attacks,” CRYPTO 2003, LNCS 2729, pp. 463-481.

[7] G. Barthe, S. Belaid, F. Dupressoir, P.-A. Fouque, B. Gregoire, P.-Y. Strub, and R. Zucchini, “Strong Non-Interference and Type-Directed Higher-Order Masking,” CCS 2016, pp. 116-129.

[8] G. Cassiers and F.-X. Standaert, “Trivially and Efficiently Composing Masked Gadgets with Probe Isolating Non-Interference,” IEEE TIFS, vol. 15, pp. 2542-2555, 2020. (CHES 2020)

[9] H. Gross, S. Mangard, and T. Korak, “Domain-Oriented Masking: Compact Masked Hardware Implementations with Arbitrary Protection Order,” TIS@CCS 2016.

[10] G. Barthe, S. Belaid, F. Dupressoir, P.-A. Fouque, B. Gregoire, and P.-Y. Strub, “Verified Proofs of Higher-Order Masking,” EUROCRYPT 2015, LNCS 9056, pp. 457-485.

[11] D. Knichel, P. Sasdrich, and A. Moradi, “SILVER – Statistical Independence and Leakage Verification,” ASIACRYPT 2020, LNCS 12491, pp. 787-816.

[12] V. Hadzic and R. Bloem, “Coco-Alma: A Versatile Masking Verifier,” FMCAD 2021.

[13] L. de Moura and S. Ullrich, “The Lean 4 Theorem Prover and Programming Language,” CADE 2021, LNCS 12699, pp. 625-635.

[14] The mathlib Community, “Mathlib4: The Lean 4 Mathematical Library,” https://github.com/leanprover-community/mathlib4.

[15] R. Iskander and K. Kirah, “From Finite Enumeration to Universal Proof: Ring-Theoretic Foundations for PQC Hardware Masking Verification,”, arXiv:2604.18717, 2026. https://github.com/rayiskander2406/qanary-universal-masking-proofs-arXiv-2604.18717.

[16] NIST, “Security Requirements for Cryptographic Modules (FIPS 140-3),” 2019.

[17] CHIPS Alliance, “Adams Bridge: Post-Quantum Cryptographic Accelerator,” https://github.com/chipsalliance/adams-bridge.

[18] NIST, “Module-Lattice-Based Digital Signature Standard (FIPS 204),” 2024.

[19] S. Faust, V. Grosso, S. Merino Del Pozo, C. Paglialonga, and F.-X. Standaert, “Composable Masking Schemes in the Presence of Physical Defaults and the Robust Probing Model,” TCHES 2018(3), pp. 89-120.

[20] R. Bloem, H. Gross, R. Iusupov, B. Konighofer, S. Mangard, and J. Winter, “Formal Verification of Masked Hardware Implementations in the Presence of Glitches,” EUROCRYPT 2018, LNCS 10821, pp. 321-353.

[21] B. Gigerl, R. Primas, and S. Mangard, “Formal Verification of Arithmetic Masking in Hardware and Software,” ACNS 2023, LNCS 13905.

[22] J.-S. Coron, J. Grossschadl, M. Tibouchi, and P. K. Vadnala, “Conversion from Arithmetic to Boolean Masking with Logarithmic Complexity,” FSE 2015, LNCS 9054, pp. 130-149.

[23] T. Tao et al., “PFR: Polynomial Freiman-Ruzsa Conjecture – Lean 4 Formalization,” https://github.com/teorth/pfr.

[24] NIST, “Transition to Post-Quantum Cryptography Standards (IR 8547),” 2025.